\documentclass[11pt]{article}

\usepackage{amssymb}
\usepackage{amsmath}
\usepackage{theorem}

\theoremstyle{plain}
\newtheorem{theorem}{Theorem}[section]

\theoremstyle{remark}

\numberwithin{equation}{section}

\begin{document}
\thispagestyle{empty}
\begin{center}
\null\vspace{-1cm}
{\footnotesize Available at: 
{\tt http://publications.ictp.it}}\hfill IC/2007/118\\
\vspace{0.5cm}
United Nations Educational, Scientific and Cultural Organization\\
and\\
International Atomic Energy Agency\\
\medskip
THE ABDUS SALAM INTERNATIONAL CENTRE FOR THEORETICAL PHYSICS\\
\vspace{1.8cm}
{\bf A SYMPLECTIC GENERALIZATION OF THE PERADZY\'NSKI\\
HELICITY THEOREM AND SOME APPLICATIONS}\\
\vspace{1.6cm}
Anatoliy K. Prykarpatsky\footnote{pryk.anat@ua.fm, prykanat@cybergal.com}\\
{\it The AGH University of Science and Technology, Krak\'{o}w 30-059, Poland,\\ 
The IAPMM of the National Academy of Sciences, Lviv, Ukraine\\
and\\
The Abdus Salam International Centre for Theoretical Physics,
Trieste, Italy,}\\[1.5em]
Nikolai N. Bogoliubov (Jr.)\footnote{nikolai\_bogolubov@hotmail.com}\\
{\it V.A. Steklov Mathematical Institute of RAN, Moscow, Russian Federation\\
and\\
The Abdus Salam International Centre for Theoretical Physics,
Trieste, Italy}\\[1.em]
and\\[1em]
Jolanta Golenia\footnote{goljols@tlen.pl}\\
{\it Department of Applied Mathematics at the AGH University of
Science and Technology,\\ Krak\'{o}w 30-059, Poland.}
\end{center}
\vspace{0.5cm}
\centerline{\bf Abstract}
\bigskip

Symplectic and symmetry analysis for studying MHD superfluid flows is
devised, a new version of the Z. Peradzy\'{n}ski \cite{Pe} helicity theorem
based on differential--geometric and group-theoretical methods is derived.
Having reanalyzed the Peradzy\'{n}ski helicity theorem within the modern
symplectic theory of differential--geometric structures on manifolds, a new
unified proof and a new generalization of this theorem for the case of
compressible MHD superfluid flow are proposed. As a by--product, a sequence
of nontrivial helicity type local and global conservation laws for the case
of incompressible superfluid flow, playing a crucial role for studying the
stability problem under suitable boundary conditions, is constructed.
\vfill
\begin{center}
MIRAMARE -- TRIESTE\\
\medskip
December 2007\\
\end{center}
\vfill

\newpage
\baselineskip=18pt

\section{Introduction}

Long ago  it was stated \cite{Ow2,Mo} that quantum vortices in superfluid
helium can be studied either as open lines with their ends terminating on
free surfaces of walls of the container or as closed curves. Nowadays  the
closed vortices are  treated  as topological objects equivalent to circles.
The existence of  structures such as knotted and linked vertex lines in the
turbulent phase is almost obvious \cite{Sc} and forces researchers to
develop new mathematical tools for their detailed investigation. In this
proposed direction it was proved by Z. Peradzy\'{n}ski \cite{Pe} a new
version of the Helicity theorem, based on some differential--geometric
methods, applied to the description of the collective motion in the
in--compressible superfluid. The Peradzy\'{n}ski helicity theorem describes
in a unique way, both the superfluid equations and the related helicity
invariants, which are, in the conservative case, very important for studying
the topological structure of vortices.

Having reanalyzed the Peradzy\'{n}ski helicity theorem within the modern
symplectic theory of differential--geometric structures on manifolds, we
propose a new unified proof and give a magneto--hydrodynamic generalization
of this theorem for the case of an incompressible superfluid flow. As a
by--product, in the conservative case we construct a sequence of nontrivial
helicity type conservation laws, which play a crucial role for studying the
stability problem of superfluid under suitable boundary conditions.

\section{ Symplectic and symmetry analysis}

We consider a quasi-neutral superfluid contained in a domain $M\subset 
\mathbb{R}^{3}$ and interacting with a ``frozen" magnetic field $%
B:M\longrightarrow \mathbb{E}^{3}$, where $\mathbb{E}^{3}:=(\mathbb{R}%
^{3},<.,.>)$ is the standard three--dimensional Euclidean vector space with
the scalar $<.,.>$ and vector ``$\times $" products. The magnetic field is
considered to be source-less and satisfying the condition $B=\nabla \times A$%
, where $A:M\longrightarrow \mathbb{E}^{3}$ is some magnetic field
potential. The corresponding electric field $E:M\longrightarrow \mathbb{E}%
^{3}$, related with the magnetic potential, satisfies the necessary
superconductivity conditions 
\begin{equation}
E+u\times B=0,\qquad \partial E/\partial t=\nabla \times B,  \label{eq2.1}
\end{equation}%
where $u:M\longrightarrow T(M)$ is the superfluid velocity.

Let $\partial M$ denote the boundary of the domain $M$. The following
boundary conditions $\langle n,u\rangle |_{\partial M}=0$ and $\langle
n,B\rangle |_{\partial M}=0$ are imposed on the superfluid flow, where $n\in
T^{\ast }(M)$ is the vector normal to the boundary $\partial M$,  considered
to be almost everywhere smooth.

Then in adiabatic magneto-hydrodynamics (MHD) quasi-neutral superfluid
motion can be described, using (\ref{eq2.1}), by the following system of
evolution equations: 
\begin{equation}
\begin{array}{c}
\partial u/\partial t=-\langle u,\nabla \rangle u-\rho ^{-1}\nabla P+\rho
^{-1}(\nabla \times B)\times B, \\[10pt]
\partial \rho /\partial t=-\langle \nabla ,\rho u\rangle ,\qquad \partial
\eta /\partial t=-\langle u,\nabla \eta \rangle ,\qquad \partial B/\partial
t=\nabla \times (u\times B),%
\end{array}
\label{eq2.2}
\end{equation}%
where $\rho :M\longrightarrow \mathbb{R}_{+}$ is the superfluid density, $%
P:M\longrightarrow \mathbb{E}^{3}$ is the internal pressure and $\eta
:M\longrightarrow \mathbb{R}$ is the specific superfluid entropy. The latter
is related to the internal MHD superfluid specific energy function $e=e(\rho
,\eta )$ owing to the first thermodynamic law: 
\begin{equation}
T\;d\eta =de(\rho ,\eta )-P\rho ^{-2}d\rho ,  \label{eq2.3}
\end{equation}%
where $T=T(\rho ,\eta )$ is the internal absolute temperature in the
superfluid. The system of evolution equations (\ref{eq2.2}) conserves the
total energy 
\begin{equation}
H:=\int_{M}\left[ \frac{1}{2\rho }|\mu |^{2}+\rho e(\rho ,\eta )+\frac{1}{2}%
|B|^{2}\right] d^{3}x,  \label{eq2.4}
\end{equation}%
called the Hamiltonian, since the dynamical system (\ref{eq2.2}) is a
Hamiltonian system on the functional manifold $\mathcal{M}:=C^{\infty
}(M;T^{\ast }(M)\times \mathbb{R}^{2}\times \mathbb{E}^{3})$ with respect to
the following \cite{HMRW} Poisson bracket: 
\begin{equation}
\begin{array}{l}
\{f,g\}:=\int_{M}\left\{ \langle \mu ,[\frac{\delta f}{\delta \mu },\frac{%
\delta g}{\delta \mu }]_{_{c}}\rangle +\rho \left( \langle \frac{\delta g}{%
\delta \mu },\nabla \frac{\delta f}{\delta \rho }\rangle -\langle \frac{%
\delta f}{\delta \mu },\nabla \frac{\delta g}{\delta \rho }\rangle \right)
\right.  \\[10pt]
\qquad \quad +\eta \langle \nabla ,(\frac{\delta g}{\delta \mu }\frac{\delta
f}{\delta \eta }-\frac{\delta f}{\delta \mu }\frac{\delta g}{\delta \eta }%
)\rangle +\langle B,[\frac{\delta g}{\delta \mu },\frac{\delta f}{\delta B}%
]_{_{c}}\rangle  \\[10pt]
\left. \qquad \quad +\langle \frac{\delta f}{\delta B},\langle B,\nabla
\rangle \frac{\delta g}{\delta \mu }\rangle -\langle \frac{\delta g}{\delta B%
},\langle B,\nabla \rangle \frac{\delta f}{\delta \mu }\rangle \right\} dx,%
\end{array}
\label{eq2.4a}
\end{equation}%
where we denoted by $\mu :=\rho u\in T^{\ast }(M)$ the specific momentum of
the superfluid motion and by $[.,.]_{_{c}}$ the canonical Lie bracket of
variational gradient vector fields: 
\begin{equation}
\lbrack \frac{\delta f}{\delta \mu },\frac{\delta g}{\delta \mu }%
]_{_{c}}:=\langle \frac{\delta f}{\delta \mu },\nabla \rangle \frac{\delta g%
}{\delta \mu }-\langle \frac{\delta g}{\delta \mu },\nabla \rangle \frac{%
\delta f}{\delta \mu }  \label{eq2.5}
\end{equation}%
for any smooth functionals $f,g\in \mathcal{D}(M)$ on the functional space $%
\mathcal{M}$. Moreover, as it was stated in~\cite{HMRW}, the Poisson bracket
(\ref{eq2.4a}) is, in reality, the canonical Lie--Poisson bracket on the
dual space to the Lie algebra $\mathcal{G}$ of the semidirect product of
vector fields on $M$ and the direct sum of functions, densities and
differential one--forms on $M$. Namely, the specific momentum $\mu =\rho
u\in T^{\ast }(M)$ is dual to vector fields, $\rho $ is dual to functions, $%
\eta $ is dual to densities and $B$ is dual to the space of two--forms on $M$%
. Thus, the set of evolution equations (\ref{eq2.2}) can be equivalently
re-written as follows: 
\begin{equation}
\begin{array}{c}
\partial u/\partial t=\{H,u\},\qquad \partial \rho /\partial t=\{H,\rho \},
\\[10pt]
\partial \eta /\partial t=\{H,\eta \},\qquad \partial B/\partial t=\{H,B\}.%
\end{array}
\label{eq2.6}
\end{equation}%
The Poisson bracket (\ref{eq2.4a}) can be re-written for any $f,g\in 
\mathcal{D}(M)$ as 
\begin{equation}
\{f,g\}=(Df,\vartheta \;Dg),  \label{eq2.7}
\end{equation}%
with $Df:=\left( \frac{\delta f}{\delta \mu },\frac{\delta f}{\delta \rho },%
\frac{\delta f}{\delta \eta },\frac{\delta f}{\delta B}\right) ^{\intercal
}\in T^{\ast }(\mathcal{M})$ and $\vartheta :T^{\ast }(\mathcal{M}%
)\longrightarrow T(\mathcal{M}),$ being the corresponding (modulo the
Casimir functionals of bracket (\ref{eq2.4a})) invertible \cite{HK}
co--symplectic operator, satisfying the standard \cite{PM,AM} properties 
\begin{equation}
\vartheta ^{\ast }=-\vartheta ,\qquad \delta (\delta w,\wedge \;\vartheta
^{-1}\delta w)=0,  \label{eq2.8}
\end{equation}%
where the differential variation complex condition $\delta ^{2}=0$ is
assumed, the differential variation vector $\delta w:=(\delta \mu ,\delta
\rho ,\delta \eta ,\delta B)^{\intercal }\in T^{\ast }(\mathcal{M})$ and the
sign ``$\ast $" denotes the  conjugate mapping with respect to the standard
bi--linear convolution $(.,.)$ of two spaces $T^{\ast }(\mathcal{M})$ and $T(%
\mathcal{M})$. Note here that the second condition of (\ref{eq2.8}) is
equivalent \cite{AM,PM} to the fact that the Poisson bracket (\ref{eq2.4a})
satisfies the Jacobi commutation condition. Thus, one can define the closed
generalized variational differential two--form on $\mathcal{M}$ 
\begin{equation}
\omega ^{(2)}:=(\delta w,\wedge \vartheta ^{-1}\;\delta w),  \label{eq2.9}
\end{equation}%
being a symplectic structure on the functional factor manifold $\mathcal{M}$
(modulo the Casimir functionals of bracket (\ref{eq2.4a})).

Denote now a subgroup $\mathcal{D}_{t}(M)=\{\varphi _{t}:M\rightarrow M\}$
of the diffeomorphism group $Diff_{+}M$, consisting of invertible
transformations $\varphi _{t}:M\rightarrow M$, generated by MHD superfluid
evolution equations (\ref{eq2.2}). This means, by definition, that 
\begin{equation}
d\varphi _{t}(x)/dt:=u(\varphi _{t}(x))  \label{eq2.10}
\end{equation}%
for all $x\in M$ and suitable $t\in \mathbb{R}$, for which solutions to (\ref%
{eq2.2}) exist and are unique. The symplectic structure (\ref{eq2.9}) is
invariant with respect to the induced mapping of diffeomorphisms $\hat{%
\varphi}_{t}:\mathcal{M}\rightarrow \mathcal{M}$ on the functional manifold $%
\mathcal{M}$, that is 
\begin{equation}
\hat{\varphi}_{t,\ast }\omega ^{(2)}=\omega ^{(2)}  \label{eq2.11}
\end{equation}%
for suitable $t\in \mathbb{R}$. Then the corresponding diffeomorphism
subgroup $\hat{\mathcal{D}}_{t}(\mathcal{M}):=\{\hat{\varphi}_{t}:\mathcal{M}%
\rightarrow \mathcal{M}\}$ satisfies the evolution equation 
\begin{equation}
d\hat{\varphi}_{t}(w)/dt:=K_{H}(\hat{\varphi}_{t}(w))  \label{eq2.12}
\end{equation}%
for any $w\in \mathcal{M}$ and the same suitable $t\in \mathbb{R}$, where
the vector field $K_{H}:\mathcal{M}\longrightarrow T(\mathcal{M})$ coincides
with the system of MHD evolution equations (\ref{eq2.2}). This fact easily
follows from the standard \cite{AM} differential--geometric considerations
related to equality (\ref{eq2.11}). Really, from (\ref{eq2.11}) one obtains
that 
\begin{equation}
0=\frac{d}{dt}\hat{\varphi}_{t,\ast }\omega ^{(2)}:=L_{K_{H}}\omega
^{(2)}=(i_{K_{H}}\delta +\delta i_{K_{H}})\omega ^{(2)}=\delta
i_{K_{H}}\omega ^{(2)}  \label{eq2.13}
\end{equation}%
for all these suitable $t\in \mathbb{R}$, where we denoted by $L_{K_{H}}$
the standard Lie derivative with respect to the vector field $K_{H}$ on $%
\mathcal{M}$ and used the corresponding Cartan formula $L_{K_{H}}=i_{K_{H}}%
\delta +\delta i_{K_{H}}$. Now, owing to the Hamiltonian equations (\ref%
{eq2.6}), the equality $i_{K_{H}}\omega ^{(2)}=-\delta H$ holds, and since $%
\delta ^{2}=0$ the invariance property (\ref{eq2.11}) is stated.

As the properties of equations (\ref{eq2.2}) on the manifold $\mathcal{M}$
are completely determined by the diffeomorphism subgroup $\mathcal{D}%
_{t}(M)\subset Diff_{+}(M)$, we will reformulate further the set of
equations (\ref{eq2.2}) making use of the suitable invariant properties on
the manifold $M$. First, observe that the mass conservation law of our
superfluid flow is equivalent to the equality 
\begin{equation}
\frac{d}{dt}\int_{D_{t}}\rho \;d^{3}x=0  \label{eq2.14}
\end{equation}%
for any domain $D_{t}\subset M$ moving together with chosen inside
particles. It is an easy calculation to rewrite (\ref{eq2.14}) in the
following equivalent form: 
\begin{equation}
\int_{D_{t}}(\partial /\partial t+L_{u})(\rho \;d^{3}x)=0  \label{eq2.15}
\end{equation}%
for all domains $D_{t}\subset M$ and suitable $t\in \mathbb{R}$, where as
above, we denoted by $L_{u}=i_{u}d+di_{u}$ the Lie derivative along the
vector field $u:M\longrightarrow T(M)$ on $M$ in the Cartan form.

As a result of (\ref{eq2.15}) one obtains the following local
differential--geometric relationship: 
\begin{equation}
(\partial /\partial t+L_{u})(\rho \;d^{3}x)=0.  \label{eq2.16}
\end{equation}
Since the evolution of our superfluid is locally adiabatic, the following
equality 
\begin{equation}
(\partial /\partial t+L_{u})\eta=0  \label{eq2.17}
\end{equation}
is obvious, meaning only that $d\eta/dt=0$ for all suitable $t\in\mathbb{R}$.

Now take the momentum conservation law in the integral Amper--Newton form 
\begin{equation}
\frac{d}{dt}\int_{D_{t}}\rho u\;d^{3}x+\int_{S_{t}=\partial
D_{t}}P\;dS_{t}-\int_{D_{t}}(j\times B)\;d^{3}x=0,  \label{eq2.18}
\end{equation}%
where $dS_{t}$ is the corresponding oriented surface measure on the boundary 
$S_{t}:=\partial D_{t}$ of a domain $D_{t}\subset M$, $P:M\longrightarrow 
\mathbb{R}$ is the internal pressure and $j:M\longrightarrow \mathbb{E}^{3}$
is the corresponding induced current density in the MHD superfluid under the
superconductivity condition. The latter means that, owing to neutrality of
the superfluid, the induction condition 
\begin{equation}
\nabla \times B+j=0  \label{eq2.19}
\end{equation}%
holds. Then from (\ref{eq2.18}) and (\ref{eq2.19}) one easily obtains the
infinitesimal form of the evolution for the velocity vector $%
u:M\longrightarrow T(M)$: 
\begin{equation}
(\partial /\partial t+L_{K_{H}})u=-\rho ^{-1}\nabla P+\rho ^{-1}(\nabla
\times B)\times B  \label{eq2.20}
\end{equation}%
coinciding, evidently, with the first equation of system (\ref{eq2.2}).

Consider now at each moment of $t\in \mathbb{R}$ the subgroup of
diffeomorphisms $\mathcal{D}_{\tau }=\left\{ \psi _{\tau }:M\rightarrow
M\right\} \subset Diff(M)$, generated by the following vector field $%
v:M\longrightarrow T(M)$ on $M$: 
\begin{equation}
d\psi _{\tau }(x)/d\tau :=v(\psi _{\tau }(x))=\rho ^{-1}B(\psi _{\tau }(x)),
\label{eq2.24}
\end{equation}
defined for a  suitable evolution parameters $\tau \in \mathbb{R}$. Since
the subgroup $\mathcal{D}_{\tau }$ does not depend explicitly on the
evolution parameter $t\in \mathbb{R},$ its action can be interpreted as
re-arranging the superfluid particles within any chosen domain $D_{t}\subset
M$. Owing now to the commutation property 
\begin{equation}
\lbrack \partial /\partial t+L_{u},L_{v}]=0,  \label{eq2.25}
\end{equation}%
equivalent to commuting subgroup $\mathcal{D}_{t}$ and $\mathcal{D}_{\tau }$
for any suitable $t,\tau \in \mathbb{R},$ from the invariance condition 
\begin{equation}
\partial \rho /\partial \tau =0,  \label{eq2.26}
\end{equation}%
we can derive that quantities 
\begin{equation}
\gamma _{n}:=L_{v}^{n}\gamma   \label{eq2.27}
\end{equation}%
for all $n\in \mathbb{Z}_{+}$ are invariants of the MHD superfluid flow (\ref%
{eq2.2}), if the density $\gamma \in \Lambda ^{3}(M)$ is also an invariant
on $M$. Really, we have 
\begin{equation}
(\partial /\partial t+L_{u})\gamma _{n}=(\partial /\partial
t+L_{u})L_{v}^{n}\gamma =L_{v}^{n}(\partial /\partial t+L_{u})\gamma =0,
\label{eq2.28}
\end{equation}%
since, by definition, we have 
\begin{equation}
(\partial /\partial t+L_{u})\gamma =0.  \label{eq2.29}
\end{equation}

Such a density can be found, observing \cite{HMRW} that the
superconductivity conditions $E+u\times B=0$, $E=-\partial A/\partial t$ and
the last equation of system (\ref{eq2.2}) brings about the invariance
condition 
\begin{equation}
(\partial /\partial t+L_{u})d\alpha ^{(1)}=0,  \label{eq2.30}
\end{equation}%
where the one--form $\alpha ^{(1)}\in \Lambda ^{1}(M)$ equals 
\begin{equation}
\alpha ^{(1)}:=\langle A,dx\rangle .  \label{eq2.31}
\end{equation}%
Moreover, since the differential operations $\partial /\partial t+L_{u}$ and 
``$d$'' commute \cite{AM}, one checks that the stronger cohomological
condition 
\begin{equation}
(\partial /\partial t+L_{u})\alpha ^{(1)}=0  \label{eq2.32}
\end{equation}%
holds on $M$, if the time--dependent gauge mapping $A\longrightarrow
A+\nabla \psi $, where $\partial \psi /\partial t+L_{u}\psi +\langle
u,A\rangle =0$, is applied to the magnetic potential $A:M\longrightarrow 
\mathbb{E}^{3}$. Now from conditions (\ref{eq2.30}) and (\ref{eq2.32}) one
easily derives  that the density 
\begin{equation}
\gamma :=\alpha ^{(1)}\wedge d\alpha ^{(1)}  \label{eq2.33}
\end{equation}%
satisfies equation (\ref{eq2.29}). Thus, it generates, in view of formula (%
\ref{eq2.27}), new conserved quantities, which can be equivalently rewritten
as 
\begin{equation}
\tilde{\gamma}_{n}:=\rho L_{v}^{n}(\rho ^{-1}\langle B,A\rangle )=\rho
L_{v}^{n}\langle v,A\rangle   \label{eq2.34}
\end{equation}%
for all $n\in \mathbb{Z}_{+}$. Thereby, the following functionals on the
functional manifold $\mathcal{M}$ 
\begin{equation}
\tilde{H}_{n}:=\int_{M}\tilde{\gamma}_{n}\;d^{3}x=\int_{M}\rho
L_{v}^{n}(\rho ^{-1}\langle B,A\rangle )\;d^{3}x  \label{eq2.35}
\end{equation}%
for all $n\in \mathbb{Z}_{+}$ are invariants of our MHD superfluid dynamical
system (\ref{eq2.2}). In particular, at $n=0$ we obtain the well-known \cite%
{HMRW} magnetic helicity invariant 
\begin{equation}
\tilde{H}_{0}=\int_{M}\langle A,\nabla \times A\rangle \;d^{3}x,
\label{eq2.36}
\end{equation}%
which exists independently of boundary conditions, imposed on the MHD
superfluid flow equations (\ref{eq2.2}).

The result obtained above can be formulated as the following theorem.

\begin{theorem}
\label{th2.1} The functionals (\ref{eq2.35}), where the Lie derivative $%
L_{v} $ is taken along the magnetic vector field $v=\rho ^{-1}B$, are global
invariants of the system of compressible MHD \ superfluid and
superconductive equations (\ref{eq2.2}).
\end{theorem}

Below we proceed to symmetry analysis of the incompressible superfluid
dynamical system and construct the related  local and global new helicity
invariants. The case of superfluid hydrodynamical flows \cite{Pu} is of
great interest for many applications owing to the very nontrivial dynamical
properties of so-called vorticity structures, featuring the motion.

\section{The incompressible superfluid: symmetry analysis and conservation
laws}

Concerning the helicity theorem result of \cite{Pe}, where the kinematic
helicity invariant 
\begin{equation}
H_{0}:=\int_{M}\langle u,\nabla \times u\rangle \;d^{3}x  \label{eq2.37}
\end{equation}%
was derived, making use of differential--geometric tools in Minkowski space
in the case of incompressible superfluid at the absent magnetic field $B=0$,
we will show below its general dynamical symmetry nature. The governing
equations look as follows: 
\begin{equation}
\partial u/\partial t=-\langle u,\nabla \rangle u+\rho ^{-1}\nabla P,\qquad
\partial \rho /\partial t+\langle u,\nabla \rho \rangle =0,\qquad \langle
\nabla ,u\rangle =0,  \label{eq2.38}
\end{equation}%
where the density conservation properties 
\begin{equation}
(\partial /\partial t+L_{u})\rho =0,\qquad (\partial /\partial
t+L_{u})d^{3}x=0  \label{eq2.39}
\end{equation}%
hold for all suitable $t\in \mathbb{R}$. Define now the vorticity vector $%
\xi :=\nabla \times u$ and find from (\ref{eq2.38}) that it satisfies the
vorticity flow equation 
\begin{equation}
\partial \xi /\partial t=\nabla \times (u\times \xi ).  \label{eq2.40}
\end{equation}%
Really, the first equation of (\ref{eq2.38}) can be rewritten as 
\begin{equation}
\partial u/\partial t=u\times (\nabla \times u)-\rho ^{-1}\nabla P-\frac{1}{2%
}\nabla |u|^{2}.  \label{eq2.41}
\end{equation}%
Then, applying  the operation ``$\nabla \times \cdot \;\;$'' \ to (\ref{eq2.41}%
), one easily obtains  the vorticity equation (\ref{eq2.40}). Moreover,
equation (\ref{eq2.40}) can be rewritten in the equivalent form 
\begin{equation}
\partial \xi /\partial t+\langle u,\nabla \rangle \xi =\langle \xi ,\nabla
\rangle u,  \label{eq2.42}
\end{equation}%
which allows a new dynamical symmetry interpretation.

Put, by definition, 
\begin{equation}
\partial x/\partial \tau =v(x,t):=\rho ^{-1}\xi ,  \label{eq2.43}
\end{equation}%
defining for all $\tau \in \mathbb{R}$ the diffeomorphism subgroup $\mathcal{%
D}_{\tau }\subset Diff\;M$ of the manifold $M$. It is easy to check that
this subgroup commutes with the previous defined subgroup $\mathcal{D}%
_{t}\subset Diff\;M$, since the following condition 
\begin{equation}
(\partial /\partial t+L_{u})v=L_{v}u  \label{eq2.43a}
\end{equation}%
holds for all $t,\tau \in \mathbb{R}$, exactly coinciding with relationship (%
\ref{eq2.42}). The condition (\ref{eq2.43a}) means that the commutation
property 
\begin{equation}
\lbrack \partial /\partial t+L_{u},L_{v}]=0,  \label{eq2.44}
\end{equation}%
similar to (\ref{eq2.25}), holds.

Now we can make use of the invariants generation technique, described above
in the case of the superfluid equations (\ref{eq2.2}). For this we need to
construct a source density invariant $\gamma\in\Lambda^{3}(M)$ of equations (%
\ref{eq2.28}) and construct successively a hierarchy of additional
invariants as 
\begin{equation}
\gamma_{n}:=L_{v}^{n}\gamma  \label{eq2.45}
\end{equation}
for all $n \in\mathbb{Z}_{+}$.

Put, by definition, $\beta^{(1)}\in \Lambda^{1}(M)$ as the one--form 
\begin{equation}
\beta^{(1)}:=\langle u, dx\rangle  \label{eq2.46}
\end{equation}
and find that 
\begin{equation}
(\partial /\partial t +L_{u})\beta^{(1)}=-\rho^{-1} dP+\frac{1}{2}%
d|u|^{2}=d(\rho^{-1}P+\frac{1}{2}|u|^{2}).  \label{eq2.47}
\end{equation}

The differential two--form $d\beta^{(1)}\in \Lambda^{2}(M) $ satisfies the
condition 
\begin{equation}
(\partial /\partial t +L_{u})d\beta^{(1)}=d^{2}(\rho^{-1}P+\frac{1}{2}%
|u|^{2})=0  \label{eq2.48}
\end{equation}
owing to the identity $d^{2}=0$. Then the differential density three--form $%
\gamma:=\beta^{(1)}\wedge d \beta^{(1)}\in \Lambda^{3}(M)$ satisfies, owing
to (\ref{eq2.47}) and (\ref{eq2.48}), the condition 
\begin{equation}
\begin{array}{c}
(\partial /\partial t +L_{u})\gamma= (\partial /\partial t
+L_{u})(\beta^{(1)}\wedge d \beta^{(1)}) \\[10pt] 
= d\left(\rho^{-1}P+\frac{1}{2}|u|^{2}\right) \wedge d
\beta^{(1)}=d\left(\left(\rho^{-1}P+\frac{1}{2}|u|^{2}\right)d
\beta^{(1)}\right).%
\end{array}
\label{eq2.49}
\end{equation}
By integration of (\ref{eq2.49}) over the whole manifold $M$ we obtain,
based on the Stokes theorem, the expression 
\begin{equation}
\begin{array}{c}
\frac{d}{dt}\int_{M}\beta^{(1)}\wedge d\beta^{(1)} =\frac{d}{dt}%
\int_{M}(u\times(\nabla\times u))\;d^{3}x=\frac{d}{dt}\int_{M}(u\times\xi)%
\;d^{3}x \\[10pt] 
=\oint_{\partial M}\left(\rho^{-1}P+\frac{1}{2}|u|^{2}\right)\langle du,
\wedge dx \rangle=0,%
\end{array}
\label{eq2.50}
\end{equation}
if the boundary conditions $\langle u, n \rangle=0$ and $\xi|_{\partial M}=0$
are imposed on the superfluid vorticity flow. Really, the surface measure $%
\langle du, \wedge dx\rangle$ on the boundary $\partial M$ can be
equivalently represented as 
\begin{equation}
\langle du, \wedge dx\rangle=\langle\langle dx,\nabla\rangle u,\wedge
dx\rangle=\langle \nabla\times u, dS\rangle=\langle\xi, dS\rangle,
\label{eq2.51}
\end{equation}
where $dS$ is the standard oriented Euclidian surface measure on $\partial M$%
. Since the vorticity vector $\xi|_{\partial M}=0$, the result (\ref{eq2.50}%
) follows automatically.

Assume now that the vorticity vector $\xi =\nabla \times u$ satisfies the
additional constraints $L_{v}^{n}\xi |_{\partial M}=0$ for $n\in \mathbb{Z}%
_{+}$. Then we obtain from (\ref{eq2.51}) and (\ref{eq2.25}) that 
\begin{equation}
\begin{array}{c}
\frac{d}{dt}\int_{M}L_{v}^{n}\gamma =\frac{d}{dt}\int_{M}L_{v}^{n}\gamma
\;d^{3}x=\int_{M}L_{v}^{n}(\partial /\partial t+L_{u})\gamma \\[10pt] 
=\int_{M}L_{v}^{n}d\left( \rho ^{-1}P+\frac{1}{2}|u|^{2}\right) d\beta
^{(1)}=\int_{M}dL_{v}^{n}\left( \left( \rho ^{-1}P+\frac{1}{2}|u|^{2}\right)
d\beta ^{(1)}\right) \\[10pt] 
=\int_{M}dL_{v}^{n}\left( \left( \rho ^{-1}P+\frac{1}{2}|u|^{2}\right)
\langle du,\wedge dx\rangle \right) =\int_{M}dL_{v}^{n}\left( \left( \rho
^{-1}P+\frac{1}{2}|u|^{2}\right) \langle \xi ,dS\rangle \right) \\[10pt] 
=\int_{\partial M}L_{v}^{n}\left( \left( \rho ^{-1}P+\frac{1}{2}%
|u|^{2}\right) \langle \xi ,dS\rangle \right) =\int_{\partial
M}\sum_{k=0}^{n}C_{n}^{k}\langle L_{v}^{k}\xi ,L_{v}^{n-k}\left( \left( \rho
^{-1}P+\frac{1}{2}|u|^{2}\right) dS\right) \rangle =0,%
\end{array}
\label{eq2.52}
\end{equation}%
bringing about the generalized helicity invariants 
\begin{equation}
H_{n}:=\int_{M}\rho L_{v}^{n}(u\times \xi )\;d^{3}x  \label{eq2.53}
\end{equation}%
for all $n\in \mathbb{Z}_{+}$. Notice here that all of the constraints
imposed above on the vorticity vector $\xi =\nabla \times u$ will be
automatically satisfied, if the condition $supp\;\xi \cap \partial
M=\emptyset $ holds. The result obtained can be formulated as the following
theorem.

\begin{theorem}
\label{th2.2} Assume that an incompressible superfluid, governed by the set
of equations (\ref{eq2.38}) in a domain $M\subset \mathbb{E}^{3}$, possesses
the vorticity vector $\xi=\nabla\times u$, which satisfies the boundary
constraints $L^{n}_{\rho^{-1}\xi}\xi|_{\partial M}$ for all $n\in\mathbb{Z}%
_{+}$. Then all functionals (\ref{eq2.53}) will be the generalized helicity
invariants of (\ref{eq2.38}).
\end{theorem}

The results obtained above allow some interesting modifications. To present
them in detail, observe that equality (\ref{eq2.47}) can be rewritten as 
\begin{equation}
(\partial /\partial t+L_{u})\beta ^{(1)}-dh=(\partial /\partial t+L_{u})%
\tilde{\beta}^{(1)}=0,  \label{eq2.54}
\end{equation}%
where, by definition, 
\begin{equation}
h:=\rho ^{-1}P+\frac{1}{2}|u|^{2},\qquad \tilde{\beta}^{(1)}:=\langle
u-\nabla \varphi ,dx\rangle ,  \label{eq2.55}
\end{equation}%
and the scalar function $\varphi :M\longrightarrow \mathbb{R}$ is chosen in
such a way that 
\begin{equation}
(\partial /\partial t+L_{u})\varphi =\nabla h.  \label{eq2.56}
\end{equation}%
Then, obviously, there holds the additional equality 
\begin{equation}
(\partial /\partial t+L_{u})d\tilde{\beta}^{(1)}=0,  \label{eq2.57}
\end{equation}%
following from the commutation property $[d,\partial /\partial t+L_{u}]=0$.
Then we obtain that the density $\tilde{\mu}:=\tilde{\beta}^{(1)}\wedge d%
\tilde{\beta}^{(1)}\in \Lambda ^{3}(M)$ satisfies the condition 
\begin{equation}
(\partial /\partial t+L_{u})\tilde{\mu}=0,  \label{eq2.58}
\end{equation}%
for all $t\in \mathbb{R}$. The similar equality holds for densities $\tilde{%
\mu}_{n}:=L_{v}^{n}\tilde{\mu}\in \Lambda ^{3}(M)$, $n\in \mathbb{Z}_{+}$: 
\begin{equation}
(\partial /\partial t+L_{u})\tilde{\mu}_{n}=0,  \label{eq2.59}
\end{equation}%
owing to the commutation property (\ref{eq2.25}). Thereby, the following
functionals on the corresponding functional manifold $\mathcal{M}$ are
invariants of the superfluid flow (\ref{eq2.28}): 
\begin{equation}
\mathfrak{M}_{n}:=\int_{M}\tilde{\mu}_{n}=\int_{D_{t}}\rho L_{\rho ^{-1}\xi
}^{n}\langle (u-\nabla \varphi ),\xi \rangle \;d^{3}x  \label{eq2.60}
\end{equation}%
for all $n\in \mathbb{Z}_{+}$ and an arbitrary domain $D_{t}\subset M$,
independent of boundary conditions, imposed on the vorticity vector $\xi
=\nabla \times u$ on $\partial M$. Notice  here that only invariants (\ref%
{eq2.60}) strongly depend on the function $\varphi :M\longrightarrow \mathbb{%
R}$, implicitly depending on the velocity vector $u\in T(M)$. Mention here
only that the practical importance of the constructed invariants (\ref%
{eq2.60}) is not still clear enough.

\section{Conclusions}

The symplectic and symmetry analysis of compressible MHD superfluid, as 
shown above, appeared to be effective for constructing the related helicity
type conservation laws, important for practical applications. In particular,
these conservative quantities play a decisive role \cite{HMRW,AK}, when
studying the stability of MHD superfluid flows under special boundary
conditions.

Here we also need to notice that the differential--geometric reformulation
of MHD equations (\ref{eq2.2}) suggested in \cite{HMRW} is incorrect.
Namely, the equality $(\partial /\partial t+L_{u})\langle \rho
^{-1}B,dx\rangle =0$ is not equivalent to the magnetic field equation $%
\partial B/\partial t-\nabla \times (\nabla \times B)=0$ that one can check
by easy calculations. Nonetheless, the commutator relation $[\partial
/\partial t+L_{u},L_{\rho ^{-1}B}]=0$ devised there and all Casimir
invariants found in article \cite{HMRW} are true. But some problems related
tothe  construction of non--Casimir type MHD superfluid flows using their
Hamiltonian structure remain, in general, open and wait still to be treated
in detail. Some of the results in this direction can be eventually obtained
making use of group--theoretical and topological tools developed in \cite%
{AK,Tr,PZ}, where the importance of the basic group of diffeomorphisms $%
Diff(M)$ of a manifold $M\subset \mathbb{R}^{3}$ and its
differential--geometric characteristics were stated.

\section*{\small Acknowledgments}

Two of the authors (N.B. and A.P.) are cordially thankful to the Abdus Salam
International Centre for Theoretical Physics in Trieste, Italy, for the
hospitality during their research 2007--scholarships. Special thanks are
tributed to Prof. J. Slawianowski (IPPT of Warsaw, Poland) for his interest
in our research, fruitful discussions and useful comments. Last but not
least thanks go to Mrs. Dilys Grilli from the ICTP Publications Office
for  her very professional  help and advice  during the preparation of the
manuscript.

\end{document}